\documentclass[prb,amsmath,twocolumn,reprint,letterpaper]{revtex4-2}
\usepackage{graphicx}
\usepackage{amssymb}
\begin{document}

\title{Density Matrix Renormalization Group study of superconducting
pairing near the quarter-filled Wigner crystal}
\author{R.~Torsten Clay}
\email{r.t.clay@msstate.edu}
\author{Beau A.~Thompson}
\affiliation{Department of Physics \& Astronomy, and HPC$^2$ Center for Computational Sciences, Mississippi State University, Mississippi State, 
 MS 39762}
 \date{today}
 \begin{abstract}
   Charge ordering is often found in the phase diagram of
   unconventional superconductors in close proximity to the
   superconducting state. This has led to the suggestion that
   fluctuations of charge order can mediate superconducting pairing.
   While several mechanisms can lead to charge order, one common
   mechanism is the long-range Coulomb interaction, resulting in a
   Wigner crystal charge ordered state.  For an electron density of
   0.5 per site we investigate the extended Hubbard model on the
   two-dimensional square lattice using exact diagonalization and
   density matrix renormalization group methods.  Our results show
   that the strength of pairing decreases with the nearest-neighbor
   Coulomb interaction strength $V$ and remains weaker than the
   tendency of pairing for non-interacting electrons.
 \end{abstract}
\maketitle

\section{Introduction}

In strongly correlated materials broken symmetry states mediated by
electron-electron (e-e) interactions are often found proximate to
unconventional superconductivity (SC).  Antiferromagnetic (AFM)
magnetic order has been studied in the most detail because of its
prominent role in the high-T$_c$ cuprates, but charge order (CO) is a
second broken symmetry state found in many unconventional
superconductors including the cuprates \cite{Frano20a}, organic
charge-transfer solids (CTS) \cite{Clay19a}, iron-based
superconductors \cite{Dagotto13a}, moire superlattice systems
\cite{Regan20a}, and other materials \cite{Neupert22a}.  As with the
many proposals that AFM fluctuations can mediate SC, similar proposals
have suggested that fluctuations of the CO can mediate pairing.

Charge order can take many forms depending on the lattice and carrier
density of the material, and can be mediated by different
mechanisms. In this paper we focus on Wigner crystal (WC) CO driven by
the long-range Coulomb repulsion.  The possibility of SC mediated by
fluctuations of the WC has been suggested for a number of different
materials with a variety of different lattice structures
\cite{Merino01a,Greco05a,Onari06a,Morohoshi08a,Kuroki06a,Watanabe06c,Tanaka04b,Nonoyama08a,Watanabe17a,Watanabe19a,Zhou11a,Yao06a}.
In this paper we focus on the simplest possible system giving WC CO,
the extended Hubbard model on a square lattice at density 0.5 per site
(quarter filling).  We present exact diagonalization and Density
Matrix Renormalization Group (DMRG) results for zero temperature
superconducting pairing correlations near the WC state. Our results
find no evidence for SC near the WC state.  The paper is organized as
follows: Section \ref{model} defines the model and correlation
functions we consider, Section \ref{results} presents our numerical
results, and Section \ref{discussion} summarizes our findings.  The
Appendix includes additional information on the implementation and
performance of the parallel DMRG method used in this work.

\section{Theoretical model and computational technique}
\label{model}

We consider the extended Hubbard model (EHM),
\begin{eqnarray}
  H &=& -t \sum_{\langle ij\rangle,\sigma}(c^\dagger_{i,\sigma}c_{j,\sigma} + H.c.)
  + U\sum_i n_{i,\uparrow}n_{i,\downarrow} \nonumber \\
  &+& V\sum_{\langle ij\rangle}n_in_j. \label{ham}
\end{eqnarray}
In Eq.~\ref{ham}, $c^\dagger_{i,\sigma}$ creates an electron of spin
$\sigma$ on site $i$; $n_{i,\sigma}=c^\dagger_{i,\sigma}c_{i,\sigma}$
and $n_i=n_{i,\uparrow}+n_{i,\downarrow}$. The sites $i$ and $j$ in
$\langle ij \rangle$ are nearest neighbor (n.n.) sites
on a square lattice.  We give energies in units
of $t$. $U$ and $V$ are the onsite and n.n.
Coulomb interactions, respectively.
We take the electron density $\rho=0.5$.

For the EHM at $\rho=0.5$ on a square lattice with $U$ large compared
to $t$, the ground state is a WC CO when $V$ exceeds a critical value
$V_c$.  At $\rho=0.5$, in one dimension (1D) the pattern of CO is
$\ldots1010\ldots$, where `0' (`1') represents a charge density
$\langle n_i\rangle$ = 0.5 - $\delta$n (0.5 + $\delta$n); in 2D
on  a square lattice
the CO
pattern is a checkerboard with $\mathbf{Q}_{\text{CO}}=(\pi,\pi)$.
$V_c$ is known exactly in 1D in the limit $U\rightarrow\infty$, where
$V_c=2$. For finite $U$, $V_c$ is larger than 2 in 1D \cite{Clay19a}.
At $U=8$, $V_c\sim 2.9$ in 1D \cite{Clay17a} and $V_c\sim 2.8$ for a
two-leg ladder \cite{Vojta01a}.  On the 2D square lattice $V_c$ is
expected to be smaller than in 1D, but is not known precisely except
in the spinless fermion limit, where on the 2D square lattice
$V_c=0.45\pm 0.02$ \cite{Song14a}.
\begin{figure}[t]
  \begin{center}
    \raisebox{9.5pc}{(a)}\hspace{0.1in}\resizebox{1.75in}{!}{\includegraphics{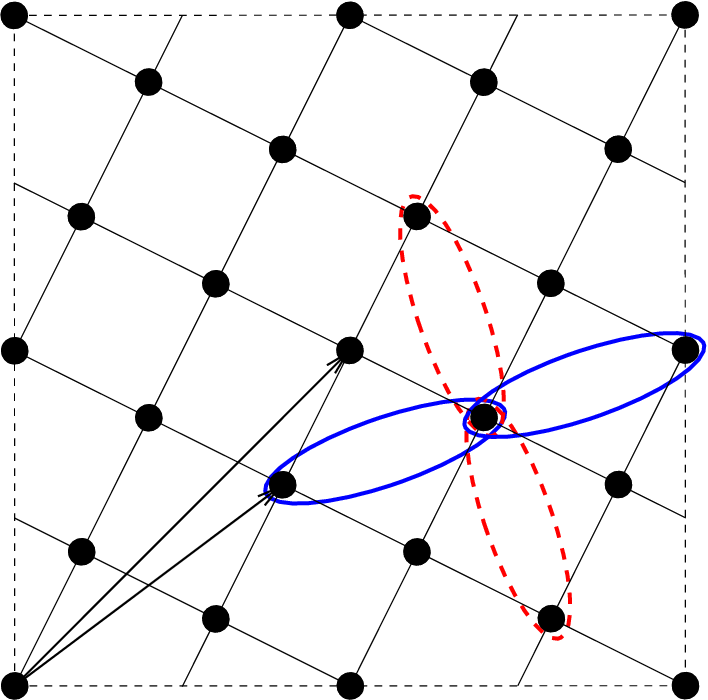}}
    \medskip
    \medskip
    
    \raisebox{8pc}{(b)}\hspace{0.1in}\resizebox{2.2in}{!}{\includegraphics{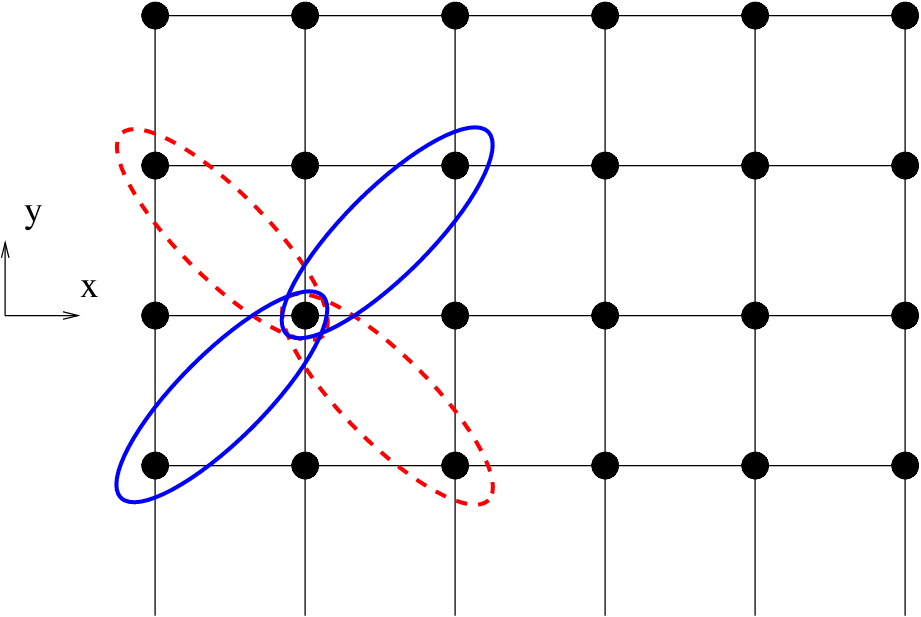}}
    \caption{(color online)  (a) Periodic 20-site
      cluster. The arrows show the furthest and next-furthest possible
      distances. (b) width four cylindrical lattice. 
        For (b) boundary conditions are periodic
      (open) in $\hat{y}$ ($\hat{x}$). A $d_{xy}$ pair is shown on each figure.
      Solid (dashed) lines correspond to singlets  with opposite sign in Eq.~\ref{delta}.   \label{lattices}}
  \end{center}
\end{figure}

There are many proposals that unconventional SC occurs in Eq.~\ref{ham}
within the metallic phase of the model with $V\lessapprox V_c$ \cite{Merino01a,Greco05a,Onari06a,Morohoshi08a,Kuroki06a,Watanabe06c,Tanaka04b,Nonoyama08a,Watanabe17a,Watanabe19a,Zhou11a,Yao06a}.
One suggestion is that following a slave boson
transformation of Eq.~\ref{ham} in the large $U$ and SU($N$) large $N$
limits, the effective bosons mediate attraction between the remaining
quasiparticles \cite{Merino01a}. Because of the strong n.n. Coulomb
repulsion, pairing would be expected to involve electrons
on next-nearest neighbor (n.n.n.) sites. While the theory was originally
applied to a square lattice subsequent work considered triangular
\cite{Tanaka04a} and other lattices \cite{Zhou11a,Yao06a}.
Calculations that support the presence of SC near Wigner crystal CO
include slave boson techniques \cite{Merino01a}, mean-field theory
\cite{Morohoshi08a}, random phase approximation (RPA)
\cite{Kuroki06a,Tanaka04a,Nonoyama08a,Zhou11a}, fluctuation exchange (FLEX)
\cite{Onari06a,Yao06a}, and variational quantum Monte Carlo (QMC)
\cite{Watanabe06c,Watanabe17a,Watanabe19a}. 

In our calculations we fix the value of $U$ at 8 and consider the
effect of increasing $V$.  Correlation functions we measure as a
function of ${\bf r}_{ij}\equiv {\bf r}_i-{\bf r}_j$ include the
charge-charge correlation $c({\bf r}_{ij})$,
\begin{equation}
  c({\bf r}_{ij}) = \langle (n_i - \langle n_i\rangle)(n_j -\langle n_j\rangle)\rangle,
  \label{cc}
\end{equation}
and the spin-spin correlation $s({\bf r}_{ij})$,
\begin{equation}
  s({\bf r}_{ij}) = \langle (n_{i,\uparrow}-n_{i,\downarrow})(n_{j,\uparrow}-n_{j,\downarrow})\rangle. \label{ss}
\end{equation}
As an order parameter of the CO phase we calculate the
charge structure factor,
\begin{equation}
  S({\bf q})=\frac{1}{N}\sum_{ij}e^{i{\bf q}\cdot{\bf r_{ij}}}c({\bf r}_{ij}),
  \label{eqsfac}
\end{equation}
at $\mathbf{Q}=(\pi,\pi)$.

We define the singlet superconducting pair creation operator as
\begin{equation}
  \Delta_i^\dagger = \mathcal{N} \sum_{v} g_v
  (c^\dagger_{i,\uparrow}c^\dagger_{i+\vec{r}_v,\downarrow}
  -c^\dagger_{i,\downarrow}c^\dagger_{i+\vec{r}_v,\uparrow}),
  \label{delta}
\end{equation}
where $\mathcal{N}$ is a normalization factor and $g_\nu=\pm 1$ a
relative phase (see Fig.~\ref{lattices}).  The pair-pair correlation
is defined as
$P(r)=\frac{1}{2}(\langle
\Delta_i^\dagger\Delta_j\rangle+\langle\Delta_i\Delta^\dagger_j\rangle)$.
For completeness we consider pairing where $\vec{r}_\nu$ in
Eq.~\ref{delta} corresponds to both n.n. and n.n.n. sites.  The most
likely pairing symmetry involves a superposition of n.n.n. pairs in
$s_{xy}$ or $d_{xy}$ form as shown in Fig.~\ref{lattices}.  In a 2D
superconductor at zero temperature, $P(r)$ must
exhibit long-range order, with
$P(r)$ extrapolating to a constant for $r\rightarrow\infty$.
We present results for two different lattices: a 20-site periodic cluster
(Fig.~\ref{lattices}(a)) which is one of the largest that can be
solved exactly, and a width four cylindrical  lattice of length $L$
with periodic boundaries in the short dimension and open
boundaries in the long dimension
(Fig.~\ref{lattices}(b)).

\section{Results}
\label{results}

\subsection{Exact diagonalization}
\label{lanczos}

While the lattice sizes available to exact solution are limited, they
can provide information on whether a given interaction strengthens or
weakens pairing correlations \cite{Clay08a}.  In particular, the
derivative of correlation functions with respect to a parameter of the
Hamiltonian (here $V$) can be very useful in locating quantum phase
transitions \cite{Clay08a}.  In Fig.~\ref{20s}(a) we plot $S(\pi,\pi)$
versus $V$.  In the thermodynamic limit a discontinuous increase in
$S(\pi,\pi)$ is expected at $V=V_c$.  The derivative
$S^\prime=dS(\pi,\pi)/dV$ should then peak at $V=V_c$. In
Fig.~\ref{20s}(a) we also plot $S^\prime$, which peaks at $V\approx
1.9$, providing an estimate for $V_c$.
\begin{figure}[tb]
  \centerline{\resizebox{3.4in}{!}{\includegraphics{fig2}}}
  \caption{Exact results for the 20 site periodic lattice
    of Fig.~\ref{lattices}(a). Solid lines are the correlation functions
    calculated on a grid of spacing $\Delta V=0.01$. Dashed lines
    are derivatives with respect to $V$ calculated using
    a centered finite difference (see text).
    (a) Charge structure factor at $\mathbf{Q}=(\pi,\pi)$
    as a function of $V$. (b)-(c) $d_{xy}$ pair-pair correlation
    function at $r=2.24$ and $r=3.16$.   \label{20s}} 
\end{figure}

For the 20 site cluster we found that 
singlet $d_{xy}$ pair-pair correlations are  the dominant
pairing symmetry; they are significantly stronger in magnitude than triplet
pair-pair correlations as well as n.n. ($s$ or $d_{x^2-y^2}$) singlet
correlations.  
In Figs.~\ref{20s}(b)-(c) we plot the $d_{xy}$ pair-pair
correlation as a function of $V$ for two values of $r$.
The values of $r$ (see Fig.~\ref{lattices}(a)) are
chosen such that the pairs in $P(r)$ do not contain overlapping lattice
points \cite{Clay08a}.
Fig.~\ref{20s}(b) shows that $P(r)$ is enhanced by $V$ for $r$=2.24, with $P(r)$
increasing with increasing $V$ compared to its value at $V=0$. However, the pair separation distance
in Fig.~\ref{20s}(b) is the shortest possible between two
non-overlapping pairs, and lattice sites comprising the pairs are
within n.n. distance of each other.  Correlations at the furthest
possible distance ($r$=3.16, Fig.~\ref{20s}(c)) only show a continuous decrease
with increasing $V$. In Fig.~\ref{20s}(b)-(c) we also plot the
derivatives $P^\prime=dP(r)/dV$ at each distance.  For a
metal--superconductor-charge order sequence of phases, $P(r)$ would
increase with $V$ and $dP/dV$ should show a peak as in
Fig.~\ref{20s}(a), and as $V$ increases further, a second negative peak at
the transition to the CO state. In the situation where no SC state
exists and instead a metal--charge order transition occurs, mobility
for both single particles as well as pairs will decrease at $V_c$,
resulting in single a negative peak in $dP/dV$ at $V_c$.  We find only a
negative peak in $dP/dV$ (Figs.~\ref{20s}(b)-(c)), suggesting the
latter situation.  The minima in $dP/dV$ is within $\Delta V\sim
\pm0.2$ of where the maxima in $dS/dV$ occurs, suggesting that both
correspond to $V=V_c$.  We do not see any evidence for a maxima in
$dP/dV$ preceding the minima, suggesting that no SC phase is
present between the metallic and CO phases.

\subsection{DMRG}

\begin{figure}
  \centerline{\resizebox{3.3in}{!}{\includegraphics{fig3}}}
  \caption{The next-nearest-neighbor pair-pair correlation function $P(r)$ for
    two different distances on a length $L=64$ cylinder
    (Fig.~\ref{lattices}(b)) versus DMRG truncation error
    $\epsilon$. Here $U=8$, $V=1$, and the maximum DMRG bond
    dimension was $m=20,000$. Solid lines show a linear extrapolation
    using the four smallest $\epsilon$ points.   \label{extrap}}
\end{figure}  

DMRG \cite{White92a,Schollwock05a} is a powerful numerical method for
studying quasi-1D quantum systems because of its accuracy and unbiased
nature. 2D systems are however challenging to solve in DMRG.
 To approach 2D,
 cylindrical lattices may be used with periodic boundaries in the
transverse direction and open boundaries in the long dimension.  In
this approach the required DMRG bond dimension $m$ increases
exponentially with respect to the transverse dimension.  Nevertheless,
DMRG results on long cylinders are useful to understand the distance
dependence of correlation functions.

Compared to the Hubbard and $t$-$J$ models, the extended Hubbard model
including $V$ has not been widely studied on 2D lattices using DMRG,
except for the case of spinless fermions \cite{Song14a,Motruk16a}.
The effect of longer-ranged Coulomb interactions has however been
studied with DMRG for some time in quasi-1D systems \cite{Anusooya97a,Sahoo12a}.
The success of the DMRG algorithm even with longer-ranged Coulomb
interactions has been attributed to the finding that diagonal
interactions do not increase entanglement entropy of states \cite{Sahoo12a}.  One
disadvantage of incorporating $V$ within DMRG is an overall slower
method due to including more interactions in the Hamiltonian.
Recently the effect
of an attractive n.n. $V$ interaction was studied on width four
cylinders of length of up to 64 \cite{Peng23a}.

Our DMRG calculations used the ITensor library \cite{itensor} and
real-space parallelization \cite{Stoudenmire13a}.  We used a two-site
DMRG update and particle number and $S_z$ conservation. To ensure
convergence of the results we performed many DMRG sweeps at small and
intermediate bond dimension and incorporated a random ``noise'' to
help prevent the method from becoming stuck in local minima. The
maximum bond dimension we used was $m=21,000$, with a minimum DMRG
truncation error of $\epsilon \sim$ 10$^{-6}$ or less.
Further details of our parallel DMRG method are given in the Appendix.

We extrapolated all observables to zero DMRG truncation error
$\epsilon$. Fig.~\ref{extrap} shows typical extrapolations of the
$d_{xy}$ pair-pair correlation function at two different distances on
the width four cylinder for the largest length we considered ($L=64$).
In the DMRG results presented in this section, the error bars on each
point are the estimated error in the linear truncation error
extrapolation. In most cases this estimated error is smaller than the
plotted symbol sizes.  Using DMRG we calculated charge and spin
correlations, pair-pair correlations for single pairs of n.n.  or
n.n.n. sites, and pair-pair correlations for $s$, $d_{x^2-y^2}$,
$s_{xy}$, and $d_{xy}$ symmetries.  On cylindrical lattices with open
boundaries, correlations of distance up to $r_{\text{max}}=L/2$ can be
measured.

\subsubsection{Charge and spin correlations}

\begin{figure}
  \centerline{\resizebox{3.2in}{!}{\includegraphics{fig4}}}
  \caption{(color online) Finite-size scaling of the charge structure
    factor for different $V$ and lattice lengths $L$. DMRG
    extrapolation errors are smaller than the symbol sizes; lines are
    fits to the points.  \label{sfac}}
\end{figure}  
\begin{figure}
  \centerline{\resizebox{3.4in}{!}{\includegraphics{fig5}}}
  \caption{(color online) (a) Charge-charge correlation function for
    the $L=64$ lattice.  (b) Charge density for the $L=64$ lattice
    with $V=1.50$. Boxes indicate local regions of competing charge order.
    (c) Bond orders.
     \label{charge}}
\end{figure}  
In open cylinders a structure factor similar to Eq.~\ref{eqsfac} can
be defined \cite{Vojta01a},
\begin{equation}
\bar{S}({\bf q})=\frac{1}{N}\sum_{ij}e^{i{\bf q}\cdot{\bf r_{ij}}}\bar{c}({\bf r}_{ij}),
\end{equation}
where $\bar{c}({\bf r}_{ij})$ is the charge-charge correlation
averaged over $N_{\rm av}$ equivalent $(i,j)$ pairs to reduce the
effect of open boundaries:
\begin{equation}
\bar{c}({\bf r}_{ij})=\frac{1}{N_{\rm av}}\sum_{(i,j)}c({\bf r}_{ij}).
\end{equation}
\begin{figure}
  \centerline{\resizebox{3.4in}{!}{\includegraphics{fig6}}}
  \caption{(color online)  Spin-spin correlation function for
    the $L=64$ lattice. Lines are fits to an exponential function (see text).
    \label{spin}}
\end{figure}  
As in Reference \cite{Vojta01a}, we take $N_{\rm av}=6$.  In the CO
phase $\bar{S}(\pi,\pi)/L$ should extrapolate to a finite value for
$L\rightarrow\infty$. Fig.~\ref{sfac} shows the finite-size scaling of
$\bar{S}(\pi,\pi)/L$, demonstrating that $V_c$ is in the range 1.50
$<$ $V_c$ $<$ 1.625.  In Fig.~\ref{charge}(a) we plot $c({\bf
  r}_{ij})$, where to mitigate boundary effects $i$ is a site on the
4th column of the lattice and $j$ a site on the same row. The large
increase in $c(r)$ between $V=1.50$ and $V=1.75$ is consistent with
the structure factor results of Fig.~\ref{sfac}. More puzzling is the
behavior of the long-range charge correlations in
Fig.~\ref{charge}(a).  For $r\alt9$ $c(r)$ increases in magnitude with
increasing $V$, but the long-distance $c(r)$ {\it decreases} in
magnitude with increasing $V$ until $V$ exceeds $V_c$. We checked
carefully that this result is not a numerical artifact by using
different DMRG starting states and by applying a pinning field to bias
the system to uniform charge order. We also verified the effect in
both $L=64$ and shorter length lattices.

The decrease in long-range charge correlation despite the increase of
$V$ can be understood by examining (see Fig.~\ref{charge}(b)-(c)) the charge density $\langle
n_j\rangle$ and bond order $\langle B_{j,j+1}\rangle$,
where
$B_{j,j+1}=\sum_{\sigma}(c^\dagger_{j,\sigma}c_{j+1,\sigma}+H.c.)$.
In $B_{j,j+1}$ we choose sites $j$ and $j+1$ as n.n. sites along the $x$
axis.  In Fig.~\ref{charge}(b) and (c) only half of the lattice is
shown, as the other half is identical due to
mirror plane symmetry. At carrier density $\rho$=0.5 charge, spin, and
bond degrees of freedom are strongly coupled, and electron-phonon
(e-p) as well as e-e interactions both play important roles
\cite{Clay19a}.  In lattices with open boundary conditions, away from
the lattice edges charge densities and bond orders take on the same
pattern found in a Hamiltonian including {\it both} e-e and e-p
interactions, in the limit of $0^+$ e-p coupling strength
\cite{Mazumdar00a}.  In 1D for $\rho$=0.5 with finite $U$ but $V<V_c$,
two density-wave states are found \cite{Ung94a}: first, a 4k$_{\rm F}$
bond order wave (BOW) state with uniform charge density and
alternating strong--weak bond orders, and second a 2k$_{\rm F}$
(period 4) charge
ordered state. These states compete with the WC CO.

As shown in Fig.~\ref{charge}(b) and (c), for $V\alt V_c$  we
find regions of the 4k$_{\rm F}$ BOW/2k$_{\rm F}$ CO state (inside
 the boxes in Fig.~\ref{charge}(b)), interspersed with regions of
$\ldots$1010$\ldots$ CO favored by large $V$ (found in between the
boxes in Fig.~\ref{charge}(b)). This shows that in
real systems with a finite e-p coupling, the phase immediately
adjacent to the $\ldots$1010$\ldots$ CO would not be metallic,
but rather 4k$_{\rm F}$ bond distorted, with possibly coexisting 2k$_{\rm F}$
charge order.
which we comment further on in Section \ref{discussion}. 

In the CO state, AFM order is expected to coexist with the CO
\cite{Clay19a}.  On a square lattice the periodicity of the AFM order
would be $\mathbf{Q}_{\text{AFM}}=(\pi,0),(0,\pi)$.  Fig.~\ref{spin}
shows the spin-spin correlations as a function of increasing $V$.  For
$V>V_c$ we do see the expected $(\pi,0),(0,\pi)$ spin order in the
short range, but spin correlations decay exponentially with distance,
indicating the presence of a spin gap.  In the CO state on the width 4
cylinder, the sites with large charge density form an effective
diagonal spin ladder (see Fig.~4 of Reference \onlinecite{Kim00a}),
which in the limit of complete CO would be expected to have a spin
gap, with spin correlations decaying exponentially with distance
($\propto e^{-r/\xi}$).  This is one difference in the open cylinders
studied here compared to a fully 2D lattice, where instead long-range
AFM order would coexist with CO.  From a fit to the spin-spin
correlations (see Fig.~\ref{spin}) we estimate $\xi\sim 2.3$ at
$V=1.75$.  For $V<V_c$ the situation is less clear. Here, we find that
the spin-spin correlations are fit slightly better by an exponential
than a power law, with for example at $V$ = 1.25 a correlation
coefficient of 0.89 (exponential fit with $\xi$ $\sim$ 6) versus 0.86
(power law fit). An exponential decay for $V<V_c$ could reflect a
small spin gap coexisting with 2 k$_{\rm F}$ CO \cite{Clay17a}, but
confirming this would likely require lattices with a larger transverse
widths.

\subsubsection{Pairing correlations}

\begin{figure}
  \centerline{\resizebox{3.4in}{!}{\includegraphics{fig7}}}
  \caption{(color online) (a) nearest-neighbor and (b)
    next-nearest-neighbor pair-pair correlations versus distance for
    the $L=32$ cylinder. The dashed lines are
    the power law $r^{-2}$.   \label{fignnn32}}
\end{figure}
\begin{figure}
  \centerline{\resizebox{3.4in}{!}{\includegraphics{fig8}}}
  \caption{(color online) (a) nearest-neighbor and (b)
    next-nearest-neighbor pair-pair correlations versus distance for
    the $L=64$ cylinder. The dashed lines are
    the power law $r^{-2}$.   \label{fignnn64}}
\end{figure}

\begin{figure}
  \centerline{\resizebox{3.4in}{!}{\includegraphics{fig9}}}
  \caption{(color online) (a) $d_{xy}$ pair-pair correlations versus
    distance for the $L=32$ cylinder. (b) $d_{xy}$
    pair-pair correlations for $L=64$.
    The dashed lines are the power law $r^{-2}$.   \label{figdxy32}}
\end{figure}

We investigated superconducting pair-pair correlation functions as a function of
$V$, $r$, and lattice size both for the correlation between individual
singlet pairs, and also for $d_{xy}$ correlations.
Fig.~\ref{fignnn32}(a) shows the correlation of singlet n.n. pairs 
along the $x$ axis
($\vec{r}_\nu=\hat{x}$ in
Eq.~\ref{delta}) for an $L$=32 lattice.  To avoid end effects we
measure pairing correlations from a pair located on the 5th/6th
lattice columns.  Because some correlations are negative we plot
$|P(r)|$.  $P(r)$ for n.n. pairs decreases rapidly with increasing
$V$, with a larger decrease upon entering the CO phase at $V=1.75$.
This is to be expected with the $V$ interaction, which suppresses
n.n. configurations in the wavefunction.  Fig.~\ref{fignnn32}(b) shows
$P(r)$ versus distance for n.n.n. singlet pairs ($\vec{r}_\nu =
\hat{x} + \hat{y}$). The n.n.n. pairing correlation also monotonically
decreases in strength with increasing $V$, a trend that is clearest in
the long-range points (10 $\leq$ $r$ $\leq$ 20 on
Fig.~\ref{fignnn32}(b)) for $V>1$. Note that the apparent increase at
$r=5$ for $V=1.5$ in Fig.~\ref{fignnn32}(b) is due to taking the
absolute value of a negative correlation.
Fig.~\ref{fignnn64} shows n.n. and n.n.n. pair-pair correlations for a
longer $L=64$ cylinder.  One difference for $L=64$ is that long range
n.n.n. correlations are slightly weaker at $V=0$ compared to $V=1$,
but like $L=32$ decrease in strength with larger $V$.
The filled circles in Fig.~\ref{fignnn32}-\ref{fignnn64} show $P(r)$
for the  uncorrelated ($U=V=0$) system. $P(r)$ for the interacting
systems is clearly weaker.

Because pairing correlations based on only single pairs may not
distinguish between $d$-wave and ``plaquette'' pairing 
\cite{Chung20a}, we also calculated $P(r)$ for
full $s$, $d_{x^2-y^2}$, $s_{xy}$, and
$d_{xy}$ pairing symmetries.  We found that pair-pair correlations for 
 $s$, $d_{x^2-y^2}$, and $s_{xy}$  pair symmetries
 were weaker than $d_{xy}$ pairing, and
show here only results for  $d_{xy}$ pair-pair correlations.
Fig.~\ref{figdxy32}(a) shows the $r$ and $V$ dependence of $P(r)$ for
$d_{xy}$ pairing on the $L=32$ cylinder.  At very short
distances ($r=3$) the $d_{xy}$ pair-pair correlation is noticeably
stronger in the interacting system.  However, again at large $r$, we
find a nearly continuous decrease of $P(r)$ with increasing $V$.  In
Fig.~\ref{figdxy32}(b) we show $P(r)$ for $L=64$, which also clearly shows
that $P(r)$ is decreases with increasing $V$ at long range.

The pairing correlations we measure appear to decay as power
laws for $V<V_c$. In a quasi-1D system of Luther-Emery type with
dominant pairing correlations such as the doped two-leg Hubbard
ladder, pair-pair correlations would however decay slower than $r^{-1}$ \cite{Dolfi15a}.
One can also argue that if pairing correlations decay slower than
$r^{-2}$ in 2D, this will lead to a diverging  susceptibility and
SC \cite{Arovas22a}.
In all cases, we find decay faster than $r^{-2}$ (see dashed
lines in Figs.\ref{fignnn32}-\ref{figdxy32}) strongly suggesting
that SC is absent in this model.

\section{Discussion and Conclusions}
\label{discussion}

We have presented charge, bond, spin, and pairing correlations for the
$\rho=0.5$ extended Hubbard model on the square lattice.  We expect
that on wider cylinders $V_c$ will be smaller than the $V_c\sim$
1.500--1.625 we find here, as is found in the spinless t-$V$ model
\cite{Song14a}. Another difference between the cylinders accessible to
DMRG versus an isotropic 2D lattice is that the ground state for
$V>V_c$ is spin gapped, with spin correlations decaying quickly with
distance as opposed to the long-range AFM order found in 2D.

We find that long-range charge correlations become weaker with
increasing $V$. This result has important implications for real
$\rho=0.5$ materials in the presence of e-p interactions,
especially the organic CTS \cite{Clay19a}. In this
case, the ground state in the $V<V_c$ region will likely not be metallic, but
will instead have another type of broken symmetry.  Calculations
including both e-e and e-p interactions on a 4$\times$4 lattice
previously found transitions between the Wigner crystal CO and other
charge/bond/spin broken symmetry states as a function of $V$
\cite{Dayal11a}.  The broken symmetry state found for $V<V_c$ depends
on several factors, principally the degree of lattice frustration
\cite{Li10a,Dayal11a}. For weak lattice frustration the 
$V<V_c$ state will have dimer-based AFM order \cite{Li10a,Dayal11a}.  For
stronger frustration, a charge-ordered Paired Electron Crystal (PEC)
occurs, with 2k$_{\rm F}$ charge order following the pattern
$\ldots$0110$\ldots$ \cite{Li10a,Dayal11a}. The spin-gapped PEC state
combines valence-bond and charge order. As $V$ increases, Fig.~\ref{charge}
shows that regions of WC CO would break up the bond-ordered
state, with the WC CO regions likely growing in size with $V$ until
long-range CO is reached at $V=V_c$.
Further DMRG calculations on
frustrated lattices will be of interest.

Our most important result is that we find no evidence for a
charge-fluctuation mediated superconducting state proximate to the WC
CO state in the EHM.  With increasing $V$, superconducting pair-pair
correlations continuously become weaker in magnitude for pair
separations beyond a few lattice spacings.  From comparing $L=64$ to
$L=32$, we do not see any signs that pairing correlations will become
dominant on longer lattices.  Furthermore, long-range pair-pair correlations in
the region near the CO state are always weaker than pair-pair correlations of
uncorrelated particles.

\begin{acknowledgments}
  This work was supported by the National Science Foundation grant number DMR-1950208.
\end{acknowledgments}

\appendix*

\section{Real space parallel DMRG}

\begin{figure}[tb]
  \centerline{\resizebox{3.2in}{!}{\includegraphics{fig10}}}
  \caption{(color online) Energy versus sweep number for the
    serial code (squares) and the real-space parallel DMRG using
    4 processors. The system was the
$L=32$ cylinder with
$U=8$ and $V=1$. From left to right, sweeps used a
    bond dimension of 3200,
  4800, 6000, 7000, 8000, 9000, and 10000.   \label{fig1app}}
\end{figure}
\begin{figure}[tb]
  \centerline{\resizebox{3.2in}{!}{\includegraphics{fig11}}}
  \caption{(color online) Energy versus DMRG truncation error
    using serial DMRG (squares)  the parallel
    algorithm  with
    four processors (triangles). System size and parameters were
    the same as Fig.~\ref{fig1app}. Lines are quadratic fits using
  the points with $\epsilon< 4\times10^{-6}$.   \label{fig2app}}
\end{figure}
In this work we used the real-space parallel DMRG algorithm presented
in Reference \cite{Stoudenmire13a}. This appendix gives further details
of our implementation of the method and its performance.

In this method, matrix-product tensors for the system are distributed
across several parallel processors, partitioning the lattice into segments of
consecutively numbered sites on each parallel process.
One advantage
of this over serial DMRG is that less memory is required on each parallel processor.
Each parallel processor
performs DMRG sweeps independently on its own lattice sites, but must
perform communications with neighboring processors in
order to update the shared bond between the partitions. Because of
this, the overall convergence of the method is somewhat slower when
compared to serial DMRG \cite{Stoudenmire13a}. In our calculations we
divided the lattice in up to $N=8$ segments. Each segment was
assigned to one node of a parallel cluster. 
Each node had two Intel Xeon 6148 processors with  20
cores per processor, and 100 Gbit/s interconnects
between nodes. For small $m$ it was usually most efficient to assign
multiple segments to one node. We also used OpenMP-based parallelization
of linear algebra within each node.

Fig.~\ref{fig1app} compares the energy computed using $N=4$ parallel
processes with serial DMRG. As expected, the convergence of the
energy versus the number of sweeps for the parallel DMRG is slightly
slower.  In practice, we find that 20-30\% more DMRG sweeps are required
using the parallel algorithm, but the increased number is offset by
the increase in speed. 
\begin{figure}[t]
  \centerline{\resizebox{3.2in}{!}{\includegraphics{fig12}}}
  \caption{(color online) Effective parallel speedup $S$ (see text) versus
    DMRG bond dimension $m$. Circles (squares) are for 4 (8) parallel
  processors. System size and parameters were
  the same as Fig.~\ref{fig1app}.   \label{fig3app}}
\end{figure}

Fig.~\ref{fig2app} compares the energy versus truncation error
for serial and parallel codes. Because of the difference in the parallel
representation of the DMRG matrix product state \cite{Stoudenmire13a}, the
truncation error for a given energy is slightly different between
the serial and parallel codes. However, we found after extrapolation
to zero truncation error, results from the serial and parallel codes agreed
closely with each other.
Fig.~\ref{fig3app} shows the effective parallel speedup $S=t_s/t_p$,
where $t_s$ is the time per sweep of the serial code, and $t_p$
the time per sweep of the parallel code. The serial code in this test was
run on one node of 40 cores using OpenMP parallelization. The
parallel code was run on four or eight nodes, with the 40 cores
within each node used for OpenMP parallelization within each lattice
partition. All parallel runs used nodes with 192 GB of memory.
The serial code was run on 192 GB memory nodes
for $m<12000$, and 384 GB of memory nodes for $m>12000$. The
serial code ran more efficiently on the large-memory nodes, which
accounts for the decrease in $S$ at $m=13000$.  Another factor
effecting the speedup was the communications latency between nodes.
While Fig.~\ref{fig3app} uses a fixed number of nodes for the
parallel calculation, we found that it was usually most efficient
to use the smallest number of nodes possible, to avoid the slower
inter- versus intra-node communications latency.
In DMRG on lattices with open boundary conditions, the
quantum entanglement is reduced for lattice sites at the boundaries.
Because of this, one parameter that must be tuned by hand to achieve the
best performance is the size of the lattice segments on each
parallel process,
which can be larger for the segments at the open boundaries
of the lattice.

Measuring non-local observables such as the pair-pair correlation
function requires computing tensor contractions across the entire
lattice. The amount of inter-processor communications to do this in
parallel would be prohibitive, so such measurements must
be performed on a single processor.  After performing the DMRG sweeps
in parallel we construct a single-processor copy of the full
wavefunction by contracting the wavefunction tensors from each node, along
with the connecting tensors (denoted $V_i$ in Reference
\onlinecite{Stoudenmire13a}). Measurements of non-local correlations
are then performed on a single processor. Because measuring correlations
takes much less memory than performing DMRG sweeps, measurements
can still be performed on a single node even when there is
insufficient memory to perform the full DMRG calculation in serial.
In practice multiple
processors can also be used to measure different observables simultaneously
to speed up the measurement process.

\end{document}